\documentclass[11pt]{article}

\usepackage[margin=1in]{geometry}
\usepackage{graphicx}
\usepackage{booktabs}
\usepackage{amsmath}
\usepackage{amssymb}
\usepackage{hyperref}
\usepackage{xcolor}
\usepackage[export]{adjustbox}
\usepackage{algorithm}
\usepackage{algpseudocode}
\usepackage{tikz}
\usetikzlibrary{arrows.meta, positioning, calc, fit}

\title{
    \vspace*{-1.5em}
    \rule{\textwidth}{0.4pt} \\[0.4em]
    \textbf{ZKLoRA: Efficient Zero-Knowledge Proofs for LoRA Verification} \\[0.4em]
    \rule{\textwidth}{0.4pt}
}
\author{ \includegraphics[height=8em]{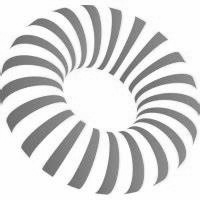} \\
    \textbf{Bidhan Roy, Peter Potash, Marcos Villagra} \\
    Bagel Research Team\footnotemark[1] \\
}
\date{\today}

\begin{document}

\maketitle

\footnotetext[1]{Bagel is a research lab, using cryptography to make open source AI monetizable.}

\begin{abstract}
  Low-Rank Adaptation (LoRA) is a widely adopted method for customizing large-scale language models. In distributed, untrusted training environments, an open source base model user may want to use LoRA weights created by an external contributor, leading to two requirements: (1) the base model user must confirm that the LoRA weights are effective when paired with the intended base model, and (2) the LoRA contributor must keep their proprietary weights private until compensation is assured.

  We present \textbf{ZKLoRA}, a zero-knowledge verification protocol that relies on succinct proofs and our novel Multi-Party Inference procedure to verify LoRA–base model compatibility without exposing LoRA weights. ZKLoRA produces \emph{deterministic} correctness guarantees and validates each LoRA module in \textbf{only 1--2 seconds} on state-of-the-art large language models. This low-latency approach enables nearly real-time verification and promotes secure collaboration among geographically decentralized teams and contract-based training pipelines. The protocol ensures that the delivered LoRA module works as claimed, safeguarding the contributor’s intellectual property while providing the base model user with verification of compatibility and lineage.
\end{abstract}

\section{Introduction}
Large Language Models (LLMs) have attained remarkable success \cite{brown2020language, devlin2018bert}, but verifying fine-tuned modifications such as LoRA \cite{hu2021lora} in an untrusted, distributed training environment can be difficult when the updated weights must remain private. 
Traditionally, one might re-run an entire forward pass or inspect thousands of parameters to ensure correctness, which is infeasible for massive models. 
\texttt{ZKLoRA} addresses this by generating a zero-knowledge proof of correctness for each LoRA module, guaranteeing that the private LoRA genuinely fits the base model. 
Crucially, \textbf{the verification stage for each LoRA module} in \texttt{ZKLoRA} remains about 1--2 seconds, even at scales of multi-billion parameter, state-of-the-art large language base models.

\section{Preliminary Results}

\begin{figure}[ht]
    \centering
    \includegraphics[width=0.9\linewidth]{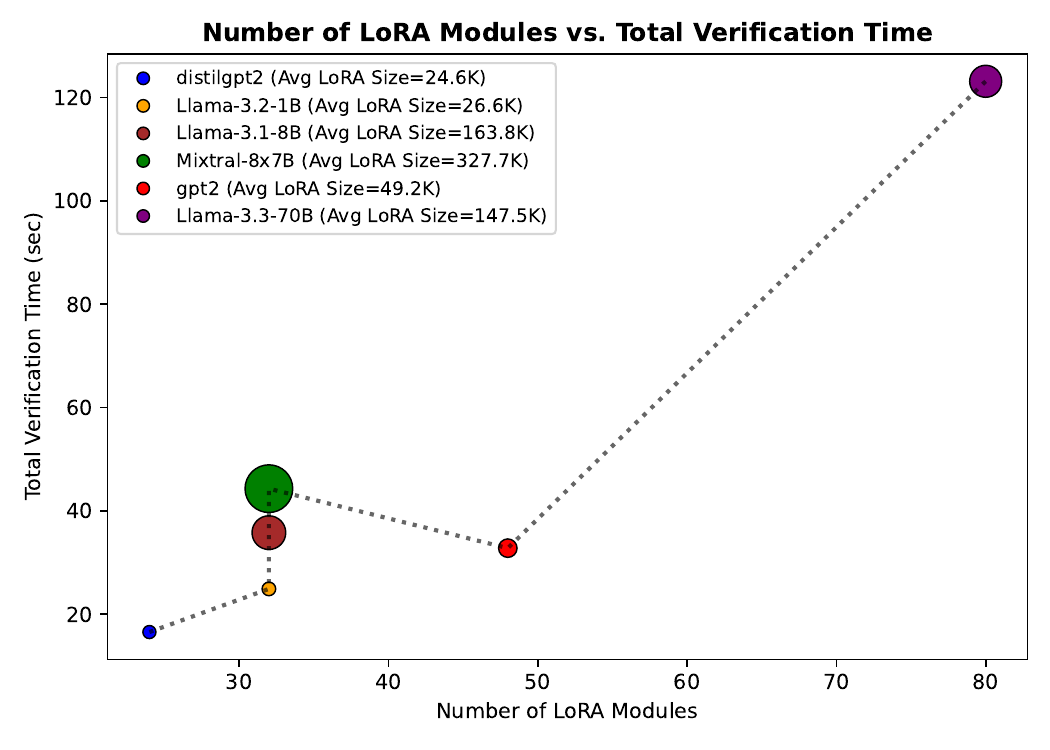}
    \caption{Total verification time (seconds) vs. number of LoRA modules, with dot size reflecting average LoRA size.}
    \label{fig:verify_vs_numlora}
\end{figure}

We benchmarked \texttt{ZKLoRA} across several LLMs and smaller models with different numbers of LoRA modules. The input for inference is a batch of size 3 with sequence length 5. Our central question is how verification times, as well as settings and proof generation times, grow with \textbf{the number of LoRA modules}, while also considering each LoRA's average parameter size. 
Figure~\ref{fig:verify_vs_numlora} and Table~\ref{tab:settings_proof} detail this trade-off.\footnotemark[2]

\footnotetext[2]{Note that the number of LoRA modules in a given model is not purely a function of the number of layers -- it is also a choice of which weight matrices within each layer is targeted. For example, just targeting one matrix per layer (ie the Query matrix in the Attention Block) will yield one LoRA per layer. Conversely, targeting the Query, Key, and Value matrices yields three matrices per layer and makes the total number of LoRAs $3 \times num$\_$layers$.}

\begin{table}[h]
\centering
\begin{tabular}{lcccc}
\hline
Base Model & \# of LoRAs & Avg LoRA Size & Avg Settings & Avg Proof \\
\hline
distilgpt2 & 24 & 24576 & 38.0 & 31.6 \\
openai-community/gpt2 & 48 & 49152 & 43.6 & 34.9 \\
meta-llama/Llama-3.2-1B & 32 & 26624 & 37.2 & 31.0 \\
meta-llama/Llama-3.3-70B-Instruct & 80 & 147456 & 54.9 & 46.9 \\
meta-llama/Llama-3.1-8B-Instruct & 32 & 163840 & 57.4 & 47.7 \\
mistralai/Mixtral-8x7B-Instruct-v0.1 & 32 & 327680 & 86.1 & 73.7 \\
\hline
\end{tabular}
\caption{Model benchmark results for Settings and Proof generation time averaged by number of LoRA modules.}\label{tab:settings_proof}
\end{table}

From Figure~\ref{fig:verify_vs_numlora}, we see that models with a higher LoRA count (e.g., 80 modules for a large 70B model) can indeed lead to larger total verification time overall. However, the slope remains modest due to \texttt{ZKLoRA}’s succinct design. For instance, even if each module is verified individually at around 1--2 seconds, verifying 80 modules can be done in just a few minutes, which is still practical for real-world usage.

Table~\ref{tab:settings_proof} similarly shows that average \emph{proof generation} and \emph{settings time} scale with the size of a LoRA module, which, combined with the number of modules, gives the total times. These two steps (proof generation on the LoRA contributor’s side and cryptographic circuit setup) can become more expensive, yet remain feasible in decentralized settings or paid contract relationships. The \emph{Base Model User}, meanwhile, benefits from the relatively short verification overhead.

Overall, these results confirm that \texttt{ZKLoRA} can handle large-scale implementation of LoRA modules with minimal overhead for verifying correctness, emphasizing the viability of repeated or multi-adapter scenarios in large-scale LLM pipelines.

\begin{figure}[ht]
\centering
\begin{tikzpicture}[
    font=\small,
    >=latex,
    thick,
    node distance=1.6cm,
    stepbox/.style={
      draw,
      rounded corners,
      minimum width=6.2cm,
      minimum height=3.2cm,
      align=center,
      fill=gray!10,
      font=\bfseries
    },
    subblock/.style={
      draw,
      rounded corners,
      align=center,
      minimum width=1.6cm,
      minimum height=0.8cm,
      fill=gray!20,
      font=\scriptsize
    },
    arrow/.style={->, thick},
    arrowlabel/.style={
      font=\footnotesize\sffamily\bfseries,
      fill=white,
      text=black,
      inner sep=1pt
    }
]

\node[stepbox] (step1) {%
  \footnotesize \textbf{Step 1: Multi-Party Inference}\\[6pt]
  \vspace{10pt}
};

\node[subblock,
      label={[font=\scriptsize]above:{Base Model User}},
      anchor=center, 
      xshift=-1.6cm,
      yshift=-1.0cm] (bmuser) at (step1.center)
{};

\node[subblock,
      label={[font=\scriptsize]above:{LoRA Contributor}},
      anchor=center, 
      xshift=+1.6cm,
      yshift=-1.0cm] (lorablk) at (step1.center)
{};

\draw[arrow] 
  ([yshift=-0.05cm]bmuser.east) 
    to[bend left=20] 
  node[midway, above, yshift=-0.1cm, font=\scriptsize]{Base Acts}
  ([yshift=-0.05cm]lorablk.west);

\draw[arrow] 
  ([yshift=-0.25cm]lorablk.west) 
    to[bend right=20] 
  node[midway, below, yshift=-0.05cm, font=\scriptsize]{LoRA Acts}
  ([yshift=-0.25cm]bmuser.east);

\node[stepbox, fill=orange!15, below=of step1] (step2) {%
  \footnotesize \textbf{Step 2: Proof Generation}\\[6pt]
  \footnotesize LoRA Contributor\\
  Builds Zero-Knowledge Proofs
};

\draw[arrow, shorten >=8pt, shorten <=8pt] (step1.south)
  -- node[arrowlabel, left, xshift=-0.1cm]{LoRA input/output acts}
  (step2.north);

\node[stepbox, fill=blue!10, below=of step2] (step3) {%
  \footnotesize \textbf{Step 3: Verification}\\[6pt]
  \footnotesize Base Model User\\
  Verifies each proof
};

\draw[arrow, shorten >=8pt, shorten <=8pt] (step2.south)
  -- node[arrowlabel, left, xshift=-0.1cm]{proof files}
  (step3.north);

\end{tikzpicture}
\caption{\textbf{Three-Step ZKLoRA Process (Vertical).}
\textbf{(1)} Base Model User and LoRA Contributor exchange “Base Acts” and “LoRA Acts” in a multi-party inference.
\textbf{(2)} The LoRA Contributor generates cryptographic proofs for correctness.
\textbf{(3)} The Base Model User verifies these proofs, ensuring correct LoRA alignment without revealing private adapter weights.}
\label{fig:zklora_3steps}
\end{figure}
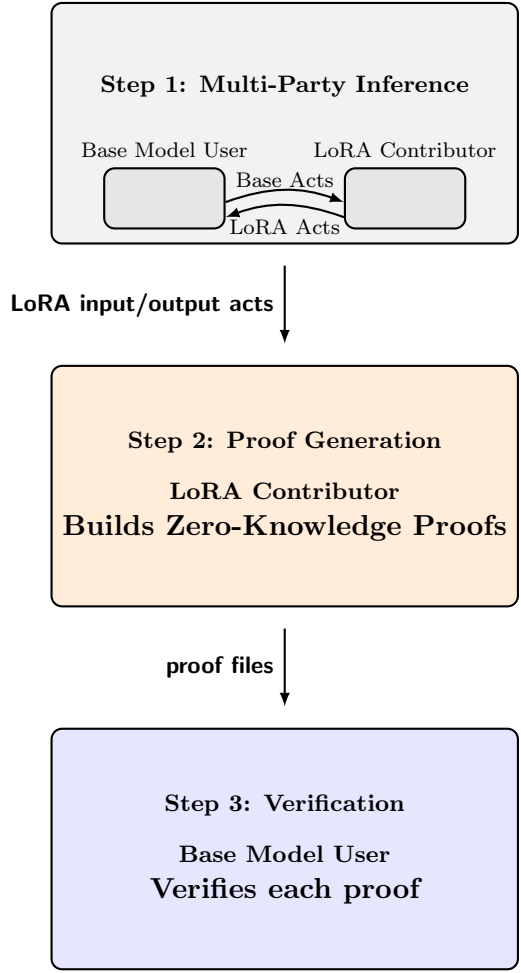

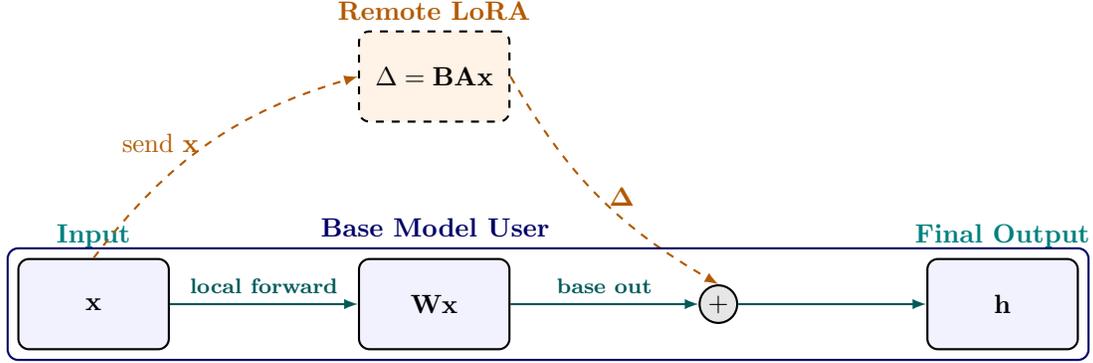
\begin{figure}[ht]
\centering
\begin{tikzpicture}[
    font=\small,
    >=latex,
    thick,
    block/.style={
      draw,
      rounded corners,
      align=center,
      minimum width=2cm,
      minimum height=1.2cm,
      fill=blue!5
    },
    dashedblock/.style={
      draw,
      dashed,
      rounded corners,
      align=center,
      fill=orange!10,
      minimum width=2cm,
      minimum height=1.2cm
    },
    sumblock/.style={
      draw,
      circle,
      fill=gray!20,
      inner sep=1pt,
      minimum size=0.5cm
    },
    arrow/.style={->, thick},
    node distance=2.5cm
]

\node[block, label=above:\textcolor{teal}{\textbf{Input}}] (xnode) {\(\mathbf{x}\)};

\node[block, right=2.5cm of xnode] (localSub) {\textbf{Wx}};

\node[sumblock, right=2.5cm of localSub] (sumNode) {\(+\)};

\node[block, right=2.5cm of sumNode, 
      label=above:\textcolor{teal}{\textbf{Final Output}}] (finalOut) {%
\(\textbf{h}\)};

\node[dashedblock, above=1.8cm of localSub, 
      label=above:\textcolor{orange!70!black}{\textbf{Remote LoRA}}] 
(lora) {\(\Delta = \mathbf{BAx}\)};

\draw[->, teal!70!black] (xnode) -- node[above, font=\scriptsize]{\textbf{local forward}} (localSub);

\draw[->, teal!70!black] (localSub) -- 
   node[above, font=\scriptsize]{\(\textbf{base out}\)} (sumNode);

\draw[->, teal!70!black] (sumNode) -- (finalOut);

\draw[->, dashed, orange!70!black] (xnode.north) to[bend left=18]
   node[left, font=\small]{send \(\mathbf{x}\)} (lora.west);

\draw[->, dashed, orange!70!black] (lora.east) to[bend right=15]
   node[right, font=\small]{\(\mathbf{\Delta}\)} (sumNode.north);

\node[
  draw,
  rounded corners,
  thick,
  color=blue!40!black,
  fit=(xnode)(localSub)(sumNode)(finalOut),
  label={[font=\small, color=blue!40!black, xshift=-1.5cm]above:\textbf{Base Model User}}
] (baseBox) {};

\end{tikzpicture}
\caption{Flow in a Multi-Party Inference scenario between local base model and remote LoRA weights. 
The base model performs a local forward pass, computing \(\mathbf{base\_out} = \mathbf{Wx}\). In parallel, the input \(\mathbf{x}\) is sent to the remote LoRA module, which returns \(\mathbf{\Delta} = \mathbf{BAx}\), where \(\mathbf{B,A}\) are the low-rank finetuned matrices. The final output is \(\mathbf{base\_out} + \mathbf{\Delta}\).}
\label{fig:remote-lora-flow}
\end{figure}

\section{ZKLoRA}
\texttt{ZKLoRA}'s design reflects the synergy between LoRA’s parameter-efficiency and zero-knowledge cryptographic protocols: LoRA significantly shrinks the parameter footprint being proven, while the zero-knowledge aspect maintains confidentiality of the contributor’s proprietary weights. By merging these ideas, \texttt{ZKLoRA} enables trust-driven collaboration across decentralized infrastructures, contract-based training, and other scenarios where proof-of-correctness is essential but the LoRA code remains private. Our approach also builds on incremental verification concepts \cite{valiant2008incrementally} and advanced proof systems such as Nova \cite{kothapalli2022nova} and HyperNova \cite{kothapalli2024hypernova}, which allow us to scale proofs to large neural networks. Ultimately, this combination provides a practical pipeline for parameter-efficient fine-tuning while verifying correctness in a succinct and minimally intrusive manner.

We implement a protocol that not only supports multi-party inference with partial activations exchanged between a base model user and a LoRA contributor, but also produces cryptographic proofs that the LoRA transforms are valid. The overall workflow is shown in Figure~\ref{fig:zklora_3steps}, while Figure~\ref{fig:remote-lora-flow} gives a deeper look at how Multi-Party Inference with LoRAs functions within an individual module. In addition, pseudocode for \texttt{ZKLoRA} is in Algorithm~\ref{alg:zklora}. To begin the Multi-Party Inference, the base model user puts the dataset chosen for inference into the base model's first module. The forward pass continues through until the base model until it hits a module that uses remote LoRA weights. When this occurs the base model user sends partial activations to the LoRA contributor for processing. These exchanged activations, shown conceptually in Figure~\ref{fig:zklora_3steps}, correspond to “Base Acts” from the base model user and “LoRA Acts” from the LoRA contributor.

After the multi-party inference finishes, the LoRA contributor shifts to a proof generation phase. At this stage, each LoRA module is compiled into a constraint system describing the LoRA transformations, and a key setup procedure yields the proving key, verification key, and possibly a structured reference string if the underlying zero-knowledge scheme requires one. The contributor then creates a “witness” by running partial activations through these constraints and finally produces the proof files. 

Once the proof generation is done, the base model user receives each proof and runs a fast verification procedure, typically requiring about 1--2 seconds per module. As Figure~\ref{fig:remote-lora-flow} suggests, this does not require the LoRA contributor to reveal the actual low-rank matrices. Instead, the contributor only sends updates and proofs that these updates conform to the declared LoRA transformations. If any single proof fails, the base model user can reject the entire LoRA submission; otherwise, the system is accepted as consistent with the underlying base model.

\begin{algorithm}[H]
\caption{\texttt{ZKLoRA} Pseudocode}
\label{alg:zklora}
\begin{algorithmic}[1]

\Require \textit{BaseModel} (public), \textit{LoRAModel} (private), \textit{Data}
\Ensure Verified outputs or \texttt{Reject}

\Statex

\State \textbf{Step 1: Multi-Party Inference}
\For{each submodule $s$ in \textit{BaseModel}}
  \If{$s$ \textbf{contains} LoRA layers}
    \State \emph{(a)} run multi-party inference with LoRA Contributor for submodule $s$
  \Else
    \State \emph{(b)} run local inference on submodule $s$ (no remote calls)
  \EndIf
\EndFor

\Statex

\State \textbf{Step 2: Proof Generation}
\For{each LoRA module $m$ in \textit{LoRAModel}}
  \State \textbf{(1) Circuit Compilation:} parse LoRA-augmented layers, produce cryptographic circuit
  \State \textbf{(2) Key Setup:} generate settings, proving key, verification key, and SRS if needed
  \State \textbf{(3) Witness Creation:} run partial activations through circuit, record wire values
  \State \textbf{(4) Proof:} construct zero-knowledge proof $\mathcal{P}_m$ for module $m$
\EndFor

\Statex

\State \textbf{Step 3: Verification}
\For{each proof $\mathcal{P}_m$}
  \If{\texttt{Verify}($\mathcal{P}_m$) fails}
    \State \textbf{return} \texttt{Reject}
  \EndIf
\EndFor

\State \textbf{return} Verified outputs

\end{algorithmic}
\end{algorithm}

\section{Related Work}

\subsection{Low-Rank Adaptation (LoRA)}
Low-Rank Adaptation (LoRA) \cite{hu2021lora} is a technique for parameter-efficient fine-tuning of large language models (LLMs) that injects small, low-rank adapter matrices into specific layers of a pre-trained model. By isolating the fine-tuning process to these low-rank components, LoRA drastically reduces memory overhead compared to full-model fine-tuning. This design choice is especially appealing for massive LLMs where training or even storing all parameters can be prohibitive \cite{ding2022delta}.

Beyond the clear advantage of reduced storage, LoRA also facilitates swapping multiple domain-specific adapters into a single base model, making it straightforward to maintain separate skill sets without instantiating an entire new copy of the model. These adapters can target specialized tasks (e.g., medical or legal text) with minimal overhead, driving LoRA’s widespread adoption. Yet verifying that a proprietary LoRA truly aligns with a base model (without revealing the adapter) remains problematic—precisely the gap \texttt{ZKLoRA} fills.

\subsection{Incrementally Verifiable Computation}
In a decentralized world, trust is a resource that is hard to achieve. In decentralized computation, we need to make sure the computations are both done and done correctly. In a seminal paper by Valiant (2008) \cite{valiant2008incrementally}, it was shown that proofs of knowledge can be used to assert the correct execution of general computations. That is, if $M$ is a machine that runs for $t$ steps producing a sequence of configurations $c_0,c_1,\dots,c_t$, then there exists an efficient and effective way to produce a computationally sound proof for the computation $c_0  \xrightarrow{t} c_t$. This idea is referred to as Incrementally Verifiable Computation or IVC.

The main goal of IVC is to produce compact, updatable proofs of correctness for a sequence of computations, so that each new step can be verified on its own while building on the guarantees of previous steps. This technique significantly reduces the verification overhead for long or evolving computations, which is invaluable in scenarios like decentralized networks, outsourced computation, and any application requiring frequent correctness checks.

Kothapalli et al. (2022) \cite{kothapalli2022nova} introduced the proof system NOVA and the idea of recursive proofs, which are proofs that can ``prove the correctness of other proofs.'' Recursive proof composition is key to IVC where each proof attests to the correctness of both a step’s output and the validity of the previous step’s proof.

HyperNova \cite{kothapalli2024hypernova} is a novel recursive argument system optimized for customizable constraint systems (CCS) that generalizes and improves upon prior approaches like Nova. It achieves efficiency through a folding scheme where the prover’s cryptographic costs are minimized and achieves zero-knowledge without relying on zkSNARKs. An IVC system allows the construction of proofs in zero-knowledge where the proofs reveal no information about the underlying computation or its inputs beyond the validity of the claim \cite{valiant2008incrementally}.

\section{Conclusion}

\texttt{ZKLoRA} provides a fast, robust mechanism to ensure that private LoRA modules remain effective when combined with a large base model. Our evaluations indicate that each LoRA module’s forward computation can be verified in less than 2 seconds, even for multi-billion-parameter LLMs. This efficiency bridges the gap between privacy-preserving LoRA development and practical, real-time validation in large-scale deployments. In terms of future work, the most relevant and immediate work would be adding polynomial commitments of the base model's activations (those that are sent as input to the LoRA Contributor). This would take us one step closer to providing end-to-end verifiability of inference computation for LoRA-finetuned models.  Other avenues of expansion could be integrating multi-contributor LoRAs, advanced zero-knowledge proofs for further performance gains, or partial data-privacy frameworks to shield user inputs as well as LoRA parameters. 

\bibliographystyle{plain}
\bibliography{references}

\end{document}